\documentclass[12pt]{article}

\begin{document} 
\topmargin 0pt 
\oddsidemargin 0mm
\renewcommand{\thefootnote}{\fnsymbol{footnote}}
\begin{titlepage}
%\begin{flushright}
%\end{flushright}                                 
\vspace{5mm}

\begin{center}
{\Large \bf Gravitomagnetic Moments and Dynamics of Dirac (spin
$\frac{1}{2}$) fermions in flat space-time Maxwellian Gravity}\\
\vspace{6mm}

{\large Harihar Behera$^a$\footnote{email: harihar@iopb.res.in} and P.C.Naik$^b$}\\
\vspace{5mm}
{\em
$^{a}$Patapur, P.O.-Endal, Jajpur-755023, Orissa, India\\
$^{b}$Department of Physics, D.D. College, Keonjhar-758001, Orissa, India}
\vspace{3mm}
\end{center}
\vspace{5mm}
\centerline{{\bf {Abstract}}}
\vspace{5mm}
The gravitational effects in the relativistic quantum mechanics are
investigated in a relativistically derived version of Heaviside's
speculative gravity ( in flat space-time) named here as ``Maxwellian
Gravity''. The standard Dirac's approach to the intrinsic spin in the
fields of Maxwellian Gravity yields the gravitomagnetic moment of a
Dirac (spin $\raise.5ex\hbox{$\scriptstyle 1$}\kern-.1em/
\kern-.15em\lower.25ex\hbox{$\scriptstyle 2$} )$ particle exactly
equal to its intrinsic spin. Violation of The Equivalence
Principle ( both at classical and Quantum-mechanical level ) in the
relativistic domain has also been reported in this work. \\
%\end{abstract}


\end{titlepage}
\section{Introduction}
The interaction of the spins of fundamental fields with gravitation is a problem of potential interest, not yet fully understood. The correct value
of the gravitomagnetic moment associated to spin \cite{1},for example, is a
question still asking for a consistent answer \cite{2}. Peres \cite
{3}  has shown that the iterated Dirac equation, in the presence of a
gravitational field, does not contain any spin-curvature coupling, in
contrast with the equation of motion of a classical spinning particle
and thereby obtained the gravitational gyro-magnetic ratio
$\,\kappa_{s}\,$ (i.e.,the gravitational analogue of the $\,g-$factor
in spin magnetic moment) of Dirac particles as zero.The authors of \cite{4}, obtained the the gravitational gyro-magnetic factor
as $\,1\,$ instead of $\,2\,$ as found in the electromagnetic
case. Many authors \cite { 2,5,6,7,8,9,10}
have found the value  $\,\kappa_{s}\,=\,g\,=\,2\,$ from other
considerations. Aldrovandi et al.\cite{5} and Wald \cite{10} define the
 gravitomagnetic moment associated to the macroscopic angular momentum
 $\vec{J}$ of a system or a body as  ${\frac{1}{2}}{\vec{J}}$,  while Mashhoon \cite{11} defines it as $\ {2}{\vec{J}}$. So it seems, there is no clear picture on the concept of gravitomagnetic moment although the problem of
  gravitational couplings of intrinsic spins of elementary particles
  is under constant analysis and have been investigated for a long time 
  both in theory and experiment ( in addition to the above
  references,see  also\cite{12,13,14,15,16,17,18} ).In \cite{16}
  Obukhov (1998) remarked that the definition and properties of a
  gravitational moment in a purely Riemannian space-time of Einstein's 
  general relativity (GR) remain unclear.
Similarly there also exists confusion,  in the
  literature, regarding the correct value of the gravitomagnetic permeability (i.e., the gravitational analogue of magnetic permeability ) that  would be the
 coefficient of a gravitomagnetic force, which is velocity dependent. This confusion stems from the following considerations.\\

The striking formal analogy between the  Coulomb's  electrostatic force
between two charges i.e $\frac{q_{1}q_{2}}{4 \pi \epsilon_0 r^2}$ 
 and the Newtonian gravitostatic force between two masses i.e. 
$ \frac{Gm_{1}m_{2}}{r^{2}} $, suggests that the analogous quantity  
for electrical permitivity $ \epsilon_{0} $ in gravitation is  
$ \epsilon_{0g} $ : 
\begin{equation}  
\epsilon_{0g} = \frac{1}{4 \pi G}   
\end{equation} 

Since any relativistic field theory of gravity would require the  existence of
finite velocity of propagation of gravitational influences i.e. gravitational waves moving at the speed of light, one can deduce  by analogy with electromagnetism ( where the speed of electromagnetic waves $ c  = {( \epsilon_{0} \mu_{0} )}^{{-1}/2} $ ), that the corresponding gravitational permitivity  and permeability are related by  $ c = {( \epsilon_{0g} \mu_{0g} )}^{{-1}/2} $. This
implies that the gravitational or gravitomagnetic permeability  must be
given by \cite{19,20} : 
\begin{equation}  
\mu_{0g} = \frac{4\pi G}{c^{2}}
\end{equation} 
and this would be the coefficient of a gravitomagnetic force, which is 
velocity dependent.\\
                                                                               General relativity (GR) predicts \cite{1,21,22} the gravitomagnetic field  $\vec B_{g, GR} $ of a spherical spinning body (such as the Earth) under slow rotation (i.e. spinning) and weak field approximation at   
\begin{equation} \vec B_{g ,GR} = 2G[\vec Jr^{2} -3( { \vec J \cdot \vec{r} }) {\cdot \vec r} ]/(c^{2}r^{5})
\end{equation}
where ${ \vec J} $ is the macroscopic spin-angular momentum of the
spinning Earth and other symbols have their usual meanings. Now, in
analogy with electromagnetism, introducing the definition of
gravitomagnetic moment  $\vec{\mu_{g}}$ of a localized mass current
distribution having angular momentum $ \vec J $  \cite{5,10}:
\begin{equation} 
\vec \mu_{g} = {\frac {\vec J}{2}}\,  ,           
\end{equation}   
Eq.(3) can be rewritten as 
\begin{equation} \vec B_{g ,GR} = 4G[\vec \mu_{g}r^{2} 
-3( { \vec \mu_{g} \cdot \vec{r} }) {\cdot \vec r} ]/(c^{2}r^{5})
\end{equation}         
Eq.(5) is formally analogous to the magnetic induction field  $ \vec B $
produced by a localized current distribution :
\begin{equation}
\vec B = -\mu_{0}[\vec \mu r^{2} - 3({\vec \mu}\cdot{\vec r})
  \cdot{\vec r} ]/(4\pi r^{5})
\end{equation} 
where  $ \mu_{0} $ is the magnetic permeability of empty space 
and  $ \vec\mu $ is the magnetic moment of the system in question. Eqs.(5) and
(6), however differ in respect of their signs. This is due to the fact 
that in electromagnetism  like  charges  repel and the unlike ones attract under static condition, but under dynamic condition the cases are
reversed,viz.,like currents (i.e.parallel currents) attract and the
unlike ( i.e. anti-parallel ) currents repel. Two magnetic poles of
the same type  interact repulsively under static condition and in dynamic condition they interact in the reversed order. In case of gravitation we encounter the opposite situation,viz, like masses interact attractively under static condition and by the nature of analogy between gravitational and electrical phenomena, we expect a reversed situation in the dynamic case, viz., like
(i.e. parallel ) mass currents should repel ( as a form of anti-gravity 
\cite{23,24} ) and the unlike (i.e. anti-parallel ) mass currents should
attract each other. Analogously the gravitational North pole-North 
pole should attract \cite{10} and the gravitational North pole-South pole  should  repel each other. This logical inference on the nature of
gravitational interaction follows naturally from general relativity.
 The deep analogy between the `gravitomagnetic moment' of a spinning
 test body in GR and the magnetic moment in electromagnetism was
 studied in \cite{10}.\\ 
 
 From the formal analogy between the Eqs.(5) and (6), one can now infer the gravitational analogue of magnetic permeability $ \mu_{0} $ in general relativity as      \begin{equation}  
\mu_{0g, GR} = \frac{16\pi G}{c^{2}} = 4\mu_{0g}
\end{equation}
which is four times the value ( Eq.(2)) expected from special
relativistic consideration. R. L. Forward \cite {23,24} used this
value ( Eq.(7)) of gravitomagnetic permeability in his discussion on the velocity dependent  forces in general relativity. In the post-Newtonian
approximation  to GR , one obtains the post-Newtonian laws of gravity
\cite {25, 26} that correspond quite closely to the Maxwellian laws of
electromagnetism, from which one can  also infer the relation  in Eq.(7). In this setting, the speed of gravitational waves  $ c_ {g,GR} $  in empty space in the weak field slow motion approximation limit of GR is expected ( in
analogy with the electromagnetic case ) at 
\begin{equation} 
c_{g, GR} = ( \epsilon_{0g}\cdot\mu_{0g, GR} )^{{-1}/2} = c/2 
\end{equation} 
i.e. half the speed of light in vacuum.This expected (or unexpected) inference of the speed of gravitational waves in GR, as in Eq.(8), can be illustrated considering the following approximations to the Maxwell-type field equations of GR in the parametrized-post-Newtonian (PPN) formalism \cite{27} (here we use
somewhat different notations for the $ \vec{g} $ and $ \vec{H} $ fields
of ref.\cite{27}) :   
\begin{equation}{\vec\nabla}\cdot\vec E_{g}\,\cong\, -4\pi G \rho_{0} 
\end{equation} 
\begin{equation}\vec\nabla\times{\vec E_{g}}\,=\,-\,\frac{1}{c}
\cdot\frac{ \partial \vec B_{g}}{ \partial t} 
\end{equation} 
\begin{equation}{\vec\nabla}\cdot{\vec B_{g}}\,=\,0 
\end{equation} 
\begin{equation}{\vec\nabla}\times{\vec B_{g}}\,=\,\left({\frac{7}{2}}\Delta_{1} + {\frac{1}{2}}\Delta_{2} \right)\left( - \frac{4\pi  G}{c}\rho_{0}\vec v + \frac{1}{c}\cdot\frac { \partial \vec E_{g}}{ \partial t}\right) 
\end{equation}
where $ \Delta_{1} $ and $ \Delta_{2} $ are PPN parameters, $ \rho_{0}
$ is the density of rest masses in the local frame of the matter, $
\vec v $ is the ordinary  (co-ordinate) velocity of the rest mass
 relative to the PPN co-ordinate frame. In general relativity  $ \left(\frac{7}{2}\Delta_{1} + \frac{1}{2}\Delta_{2}\right)\,\cong\, 4 $  and so Eq.(12) can be rewritten as
\begin{equation}{\vec\nabla}\times{\vec B_{g}}\,\cong\,-\,\frac{16\pi  G}{c}\rho_{0}\vec v + \frac{4}{c}\cdot\frac { \partial \vec E_{g}}{ \partial t} 
\end{equation} 
In empty space (where $ \rho_{0} = 0 $), these field equations reduce 
to the following equations: 
\begin{equation}{\vec\nabla}\cdot\vec E_{g} = 0 
\end{equation}
\begin{equation}{\vec\nabla}\times{\vec E_{g}} = -\,\frac{1}{c}\cdot\frac{ \partial\vec B_{g}}{ \partial t} 
\end{equation}
\begin{equation}{\vec\nabla}\cdot{\vec B_{g}} = 0
\end{equation}                                                                 \begin{equation}{\vec\nabla}\times{\vec B_{g}} =  \frac{4}{c}\cdot\frac { \partial \vec E_{g}}{ \partial t} 
\end{equation}
Now taking the curl of (15) and utilizing Eqs.(14) and (17) we get the wave equation for the field  $ \vec E_{g} $ in empty space as 
\begin{equation}{\vec\nabla^{2}}\cdot\vec E_{g} -  \frac{1}{{c_{g}}^{2}}\cdot{\frac{\partial^{2}\vec E_{g}}{\partial t^{2}}}  = 0 
\end{equation}
where  $ c_{g} = c/2 $. Similarly the wave equation for the field  $
\vec B_{g} $ can be obtained  by taking the curl of Eq.(17) and
utilizing Eqs.(15) and (16) : 
\begin{equation}{\vec\nabla^ {2}}\cdot\vec B_{g} - \frac{1}{{c_{g}}^{2}}\cdot\frac{\partial^{2}\vec B_{g}}{\partial t^{2}} = 0
\end{equation} 
where  again we get  $ c_{g} $ = c/2 . This is against the special relativistic
 ( as well as the gauge field theoretic ) expectation that the speed of gravitational waves (if they exist) should be equal to the speed of light in any Lorentz-covariant field theory of gravity. It is to be noted that the factor of 4 is
responsible for this result and we shall come to the
question of the origin  of this factor of 4 later on.\\ 
      
 In this paper an attempt is made to get some  better understanding of
 these  problems among others within the framework of a
 relativistically derived version of Heaviside's \cite{28} speculative
 gravity named here as ``Maxwellian Gravity''.This theory is a vector
 model theory of gravity in  Minkowski space-time having
 striking similarity as well as characteristic dissimilarity ( due to
 the observed  attractive nature  of gravity) with  Maxwell's
 electromagnetic theory. We named it as ``Maxwellian gravity''
 because as far as we are aware, Maxwell first attempted to develop a
 vector theory of gravity, but he left it because some problem arose with which he became dissatisfied. We will discuss and attempt to resolve the  problem faced by Maxwell later in this paper. Heaviside,  pursued
 further Maxwell's left  work and developed the full set of
 Lorentz-Maxwell-type  field equations for gravity by virtue of his
 power of  speculative thought. Unfortunately he did not work further
 on it for reasons unknown to us and his theory  has not been
studied as thoroughly as it deserves.
 Regrettably,the standard texts on gravitation do not
contain any reference to Heaviside's work, but we came to know about
his work  from McDonald \cite{28} who described Heaviside's gravity as 
a low velocity and weak field approximation to General
Relativity. However,in this work, we shall see how the gravitational
equations speculated by Heaviside can now be derived with ( and also
without ) the aid of special relativity  and see what insights come out of these  equations regarding our understanding
of the physical world. To attain these  objectives among others, it is thought necessary to begin with a resume of the development of the field equations of Maxwellian Gravity and some of their important and immediate consequences  in the following two consecutive sections before coming to the topic of this paper which is described in the subsequent sections.\\
       
\section{Maxwellian Gravity} 
Maxwellian Gravity is a Faraday-Maxwell-type field theory of gravity in Minkowski space-time. In its formulation, the relativistic nature of
gravity and the source of gravity are explored in
the following line of thought.\\ 
 
 It is well known that Newton's action-at-a distance  theory of gravitation is incompatible with special relativity (SR) because it violates the principle of causality by suggesting gravitational influences propagating through space at an infinite speed. In Newton's gravitation theory, the meaning of  ``mass'' i.e the gravitational mass \cite{29,30,31}, became ambiguous with the establishment of SR because SR suggests two distinct mass concepts,viz.,one Lorentz-invariant rest mass \cite{32} and other velocity dependent (i.e. frame dependent ) inertial mass ( better say  relativistic mass ) which is not Lorentz-invariant.  In this setting, one key question arises,viz., what form of mass (or  energy ) represents the gravitational mass ? Thus in order to construct a field theory of gravity compatible with SR and the correspondence principle by which  the  relativistic gravity is  reducible  to Newtonian gravity, a decision on what ``mass'' (or energy) of SR  represents the gravitational  mass, has to be taken. Such a decision,  Price \cite{33} has rightly pointed out, will be crucial not only to the  resolution of the ambiguity already mentioned but also to the issue of the non-liner nature of gravity.\\
In his formulation of GR ,Einstein has taken a decision  in favor of
the equivalence of inertial and gravitational masses which he
expressed \cite{34,35} as : \begin{quote}``The proportionality between
  inertial and gravitational masses holds for all bodies without
  exception, with the [experimental] accuracy achieved thus far, so
  that we may assume its general validity until proved
  otherwise''. \end{quote}   Although  Einstein's assumption of the proportionality of gravitational and inertial masses came out of the experimental evidence available at that time but later on it could be verified to different degrees of accuracy in numerous experiments \cite{1,29}. However, it is noted by Mashhoon \cite{36} in 1993 that the observational evidence for the principle of equivalence of gravitational and inertial masses is not yet precise enough to reflect  the wave nature of matter and radiation in their interactions with
gravity \cite{37,38,39,40}. The equivalence of gravitational and
inertial masses is the basis for Einstein's principle of equivalence
between a gravitational field and an accelerated frame of reference
\cite{34,35,36}. It is to be noted that authors like Fock \cite{41}
and M{\o}ller \cite{42} disagree  about the equivalence of accelerated frames (or observers) and gravitational fields, while Synge \cite{43} does not seem to believe
that there is such a principle at all. The doubts generated by Fock
and Synge have not disappeared completely and the equivalence principle 
 has attracted considerable attention in the past decades; see the
 review \cite{44}.Recently the validity of the equivalence principle
 has also been questioned or suspected at the quantum level by many
 authors,see for example\cite{12,13,17,18,45,46}. Hammond \cite{46} at
 one point, in some vein, noted : ``the only thing the principle of equivalence proves is that it is wrong''.The violation of the equivalence has also been examined  in the literature as a possible solution  to the solar neutrino problem ( see \cite{47,48} and  the literature quoted therein ). In view these controversies over the equivalence principle, let us look back closely at the theoretical logic behind Einstein's assumption of the equality  of gravitational and inertial masses. Einstein \cite{31},by writing out the Newton's equation of  motion in a gravitational field,in full :
\begin{equation}\rm {(Inert\,\, mass)(Acceleration) = (Gravitational\,\,  mass)(Intensity\,\,of\,\, gravitational\,\,field)} 
\end{equation} 
inferred from it:{\it `` It is only when there is numerical equality between the inert and gravitational mass  that the acceleration is independent of
the nature of the body''}. This inference is often expressed in one of
two ways :       \\
$(A_{1}){\it\, {that\,\, the\,\, motion\,\, of\,\, the\,\, particle\,\,
    is\,\, mass\,\, independent,\,\,\,\, or}}      \\
(A_{2}){\it\, {that\,\, the\,\, inert\,\, mass\,\, of\,\, the\,\, particle\,\, is\,\, equal\,\, to\,\, its\,\, gravitational\,\,
mass.}}$        \\
  The two statements $ (A_{1})\,\, and\,\, (A_{2})$ are sometimes used
 interchangeably as the {\it weak equivalence principle} ({WEP}) in 
 the literature \cite{1,2,29,30}. This use of terminology is rather
 confusing, as the two statements are logically independent
 \cite{45}. They happen to coincide in the context of
 Galileo-Newtonian physics but may diverge in other settings. This is
 what may happen in the relativistic case as we shall see in this
 work. The expectation of such a possible divergence of the statements
 $ (A_{1})\,\,and\,\,(A_{2}) $ stems from the observation that the
 above cited inference of Einstein is non-relativistic in the sense
 that  it  is drawn from a  non-relativistic equation (20) and there
 exists the Lorentz-invariant  mass (or energy) that might  possibly  be treated as the gravitational analogue of the Lorentz-invariant electric charge in  developing a  Lorentz-covariant theory of gravity. To explore and illustrate the possibility of identifying the rest mass as the gravitational analogue of the electric charge, to get new insights for  developing a special relativistic generalization of Newtonian gravity, to regard old problems from a new angle, let us make a re-investigation of an often  cited \cite{20,36} thought experiment\cite {49} from a new angle as under.\\ 
        
    Consider a system of two non-spinning massive point-like charged
 particles having such amount rest masses  $ m_{01}\,\,and\,\,m_{02}
 $, with respective electric charges  $ q_{1}\,\,and\,\,q_{2} $  that
 they are at rest in an inertial frame  $ K^{\prime} $, under 
 equilibrium condition due to a mutual balance of the force of
 Coulombic repulsion and the Newtonian gravitostatic attraction
 between them. Our task is to investigate the condition of equilibrium
 the said particle system in different inertial frames in relative
 motion. To this end, suppose that the particles are positively charged 
 and they are in empty space. Let the particle No.2 be positioned at the
 origin of  $ K^{\prime} $-frame and $ \vec r_{0} $ be the position
 vector of the No.1 with respect to the particle No.2. In this frame
 the condition of equilibrium may be represented as 
\begin{equation}\vec F_{C} + \vec F_{N} = \frac{q_{1} q_{2} \vec
    r_{0}}{4\pi \epsilon_{0}{r_{0}}^{3}} - \frac{G m_{01} m_{02} \vec
    r_{0}}{{r_{0}}^{3}} = 0    \end{equation}
where $\,\vec F_{C}\,$ and $\,\vec F_{N}\,$ respectively denotes the
Coulomb and Newton force and the other symbols have their usual meanings. From Eq.(21) we get
\begin{equation}q_{1} q_{2} = 4\pi \epsilon_{0} G m_{01} m_{02}
\end{equation}     
Eq.(22) represents the condition of equilibrium, in terms of the
charges and masses of the particles, under which an equilibrium will
be effected in the $ K^{\prime} $-frame.  \\ 
       
Now, in order to investigate the problem of equilibrium of the said
particle system from the point of view of an observer in another
inertial frame $ K $ in uniform relative motion with respect to the $
K^{\prime} $-frame and to simplify the investigation , let the
relative velocity  $ \vec v $ of $ K $ and $ K^{\prime} $-frames be
along a common  $ X /X^{\prime}- $ axis with corresponding planes
parallel as usual. Since the particles are at rest in $ K^{\prime}
$-frame, both of them have the same uniform velocity $ \vec v $
relative to the $ K $-frame. Let the position vector of the particle
No.1 with respect to the particle No.2 as observed in $ K $-frame be $
\vec r $ and the angle between  $ \vec v  $ and  $ \vec r $  be  $
\theta $.\\ 
                                                                               For an observer in  $ K  $-frame, the force of electric origin on either particle (say on particle No.1 due to the No.2 particle ) is no more simply a Coulombic force, but a Lorentz force, viz., 
\begin{equation}\vec F_{L}  =  q_{1}\,\vec E_{2} +\, q_{1}{\vec v}\times{\vec B_{2}}
\end{equation}
 where 
\begin{equation}\vec E_{2}\,\,=\,\,\frac {q_{2}\, (\, 1 - v^{2} /
    c^{2}\, )\,\,\vec r}{ 4 \pi \epsilon_{0}\, r^{3}\, {[\,\, 1 - {( v^{2}/c^{2})\,\, {sin}^{2} \theta}\,\,]}^{3/2}}         \end{equation} 
\begin{equation}\vec B_{2}  =  \frac { \vec v \times \vec E_{2}}{c^{2}} 
 \end{equation} 
\begin{equation}\vec r  =  \frac { \vec r_{0}\,\,[\,\,1 - ( v^{2}/c^{2})\,\,{sin}^{2} \theta\,\,]^{1/2}}{{[\,\, 1 - ( v^{2} / c^{2})\,\,]}^{1/2}} 
\end{equation} 
and the symbols have their usual meanings.  \\ 
      What about the force of gravitational interaction as observed in
      the $ K $-frame ? It can not simply be a Newtonian force but
      something else, otherwise the particle system will not remain in 
      equilibrium in the $ K $-frame. Such a situation will amount to
      a violation of the relativity principle of special
      relativity. Therefore  a new force law of gravity has to be
      invoked so that the equilibrium is maintained in accordance with
      the relativity principle. Let this new force be represented by
      $ \vec F_{gL} $ such that  the equilibrium condition in $ K
      $-frame be satisfied as :  \\ 
\begin{equation}\vec F_{gL}\,\,+\,\,\vec F_{L}\,\,= 0
\end{equation}
Taking into account the Eqs.(23)-(26), $ \vec F_{gL} $ in Eq.(27) can be
expressed  as :  
\begin{equation}\vec F_{gL} = -\vec F_{L}\,=\,-\,\frac {q_{1}\,q_{2}\,(\,1\,-\,v^{2}/ c^{2}\,)^{2}\,\,\vec r}{4 \pi \epsilon_{0}\,\,r^{3}\,\,[\,1\,-\,\,(v^{2}/c^{2})\,\,{\sin}^{2}\theta\,]^{3/2}}\,\,-\,\,\frac{ q_{1}\, q_{2}\,(\vec v\cdot\vec r)(\, 1 - v^{2} / c^{2}\, )\,\,\vec v}{ 4 \pi \epsilon_{0}\, r^{3}\, {[\,\, 1 - {( v^{2}/c^{2})\,\, {sin}^{2} \theta}\,\,]}^{3/2}} 
\end{equation} 
Now with the help of Eq.(22),we can eliminate $\,q_{1}\,q_{2}\,$ from
Eq.(28) and get the expression for $\,\vec F_{gL}\,$ as : 
\begin{equation}\vec F_{gL}\,=\,-\,\frac {G\,m_{01}\,m_{02}\,(\,1\,-\,v^{2}/ c^{2}\,)^{2}\,\,\vec r}{\,r^{3}\,\,[\,1\,-\,\,(v^{2}/c^{2})\,\,{\sin}^{2}\theta\,]^{3/2}}\,\,-\,\,\frac{\,G\,m_{01}\,m_{02}\,(\vec v\cdot\vec r)(\, 1 - v^{2} / c^{2}\, )\,\,\vec v}{ c^{2}\, r^{3}\, {[\,\, 1 - {( v^{2}/c^{2})\,\, {sin}^{2} \theta}\,\,]}^{3/2}} 
\end{equation} 
Eq.(29) may be rearranged into the following form :                            \begin{equation}\vec F_{gL}  =  m_{01}\,\vec E_{g2} +\, m_{01}{\vec v}\times{\vec B_{g2}}
\end{equation}
 where 
\begin{equation}\vec E_{g2}\,\,=\,-\,\frac {G\,m_{02}\, (\, 1 - v^{2} /
    c^{2}\, )\,\,\vec r}{\, r^{3}\, {[\,\, 1 - {( v^{2}/c^{2})\,\, {sin}^{2} \theta}\,\,]}^{3/2}}         \end{equation} 
\begin{equation}\vec B_{g2}\, =\,\frac { \vec v \times \vec E_{g2}}{c^{2}} 
\end{equation}
Eqs.(30)-(32) are in complete formal analogy with Eqs.(23)-(25) of
classical electromagnetism in its relativistic version. Thus, from the
requirement of the frame-independence of the equilibrium conditions,
we not only obtained a gravitational analogue of the  Lorentz-force law
 (or the gravitational Lorentz force law in Eq.(30)) but also unexpectedly found the Lorentz-invariant rest mass as the gravitational analogue of the
electric charge by analogy. From this analysis, the gravitational
charge (or mass) invariance may be interpreted as a consequence of the
Lorentz-invariance of the physical laws. These findings are in
conformity with Poincar\'{e}'s \cite {50} remark that {\em if equilibrium is to be a frame-independent condition, it is necessary for all forces of non-electromagnetic origin to have precisely the same transformation law as that of the Lorentz-force}. Having recognized these findings we  obtained four Faraday-Maxwell-type linear equations of gravity describing what we call``Maxwellian Gravity'' following the known procedures of the electromagnetic theory; see for example, an excellent text by Rosser\cite{51}. The resulting equations have a surprisingly rich and detailed correspondence with Faraday-Maxwell's field equations of the electro-magnetic theory. The field equations can be written in the following Faraday-Maxwellian-form:\\
\begin{equation}
\vec{\nabla}\cdot{\vec{E}_{g}} = -4\pi G \rho_{0} =
-{\rho_{0}}/{\epsilon_{0g}}
\; \;\;\;\;\;\; \; \; {by\,\,\, defining \;\;\;\;  \epsilon_{0g}={1}/4\pi G}
\end{equation}
\begin{equation}
\vec{\nabla} \times \vec{B}_g = - \mu_{0g} \vec{j}_0 + (1/c^2)
({\partial \vec{E}_g }/{\partial t}) ,\;\;\;\;by\,\,\, defining\;\;\;\; \mu_{0 g}
= {4{\pi}G}/{c^{2}}
\end{equation}
\begin{equation}
\vec{\nabla} \cdot \vec{B}_{g} = 0
\end{equation}
\begin{equation}
\vec{\nabla} \times \vec{E}_g = - \partial \vec{B}_g / \partial t
\end{equation}
Where $\rho_{0}$ = rest mass (or proper mass)  density; $\vec{j_{0}}$
= rest mass current density; $G$ is Newton's universal gravitational
constant; $ c $ is the speed of light in empty space; the gravito-electric and gravito-magnetic
fields $\vec{E_{g}}$ and $\vec{B_{g}}$  respectively are defined by
the gravitational Lorentz force on a test particle of rest mass $
m_{0} $ moving with uniform velocity $ \vec{u} $ as
\begin{equation}
\frac{d}{dt}[m_{0}\vec{u}/(1-{u^2}/{c^2})^{1/2}] = m_{0}[\vec{E_{g}}+\vec{u}\times\vec{B_{g}}]
\end{equation}
where the symbols have their respective meanings in correspondence with
the Lorentz force law in its relativistic form. Interestingly the
field equations (33)-(36) happen to coincide structurally with those
speculated by Heaviside \cite{28}.
In covariant formulation, introducing the space-time four vector
$x_{\mu} = (x,y,z,ict)$, proper
mass current density four vector $j_{\mu}=(j_{0x},
j_{0y},j_{0z},ic\rho_{0})$  and the second-rank antisymmetric
gravitational field strength tensor   
\begin{eqnarray}
F_{\mu\nu} = \pmatrix {
0 & B_{gz} & -B_{gy} & -iE_{gx}/c\cr
-B_{gz} & 0 & B_{gx} & -iE_{gy}/c \cr
B_{gy} & -B_{gx} & 0 & -E_{gz}/c \cr
iE_{gx}/c & iE_{gy}/c & iE_{gz}/c & 0}
\end{eqnarray}                 
The field equations (33-36) can now be represented by the following two
equations:
\begin{equation}
\sum_{\nu}{\partial F_{\mu \nu}}/{\partial x_{\nu}} = -\mu_{0g}{\vec j_{\mu}}, \; \;  {\rm where} \; \; \mu_{0g} = 4{\pi}G/{c^{2}}
\end{equation}
\begin{equation}
\partial F_{\mu \nu}/ \partial x_{\lambda} +\partial F_{\nu \lambda}/ {\partial x_{\mu}} + \partial F_{\lambda \mu}/ {\partial x_{\nu}} = 0
\end{equation}
while the gravitational Lorentz force law (37) assumes the form :
\begin{eqnarray}
c^{2}(d^2 x_{\mu}/ds^2) = F_{\mu \nu}({dx^{\nu}}/ds)
\end{eqnarray}
The absence of the rest mass of the test particle in its co-variant equation
of motion (41) in the external gravitational field $ F_{\mu\nu} $
describes clearly the universality of free fall  or  the uniqueness of
free fall (UFF) known since the time of Galilei. Here in the relativistic
domain we saw the UFF also holds true for non-spinning particle-field
system but in addition to this  we also  notice here  a divergence of the statements $\,(A_{1})\,\,and \,\,(A_{2})\, $ stated earlier after Einstein's
inference from Eq.(20).\\ 
   
The present analysis seems to establish an equivalence of gravitational mass and rest mass ( or any form of Lorentz-invariant mass-energy) and predict the existence of a magnetic-type component in gravity. Although the inertial mass of a body depends on its  total energy content as revealed by special relativity, this mass seems not to represent the gravitational charge (or mass ) as per the revelations of this analysis which differs from that made by the original designers of the thought experiment\cite {47}, who axiomatically used the inertial mass as the gravitational charge (or mass). However,the present derivation
of the gravitomagnetic field agrees with  the works of Sciama
\cite{52} and Bedford and Krumm\cite {53} who axiomatically used the
rest mass as the gravitational mass. It is to be noted that the main
purpose of this analysis of the thought experiment is to illustrate
how the rest mass of a particle manifests itself as the  gravitational
analogue of the electric charge in a specific situation. Since  the
rest mass is seen here as playing the role of the gravitational charge
(or mass) in a specific situation, we have reasons to expect it to
play  the same role in other situations also even when the electric
force is not balanced by a gravitational force. By assuming the equality of gravitational and rest masses, the validity of Newton's law of gravitation and the
principle of Lorentz-invariance of all physical laws, the
gravitational analogues of Maxwell-Lorentz equations (33)-(37) ,can
also be obtained following the methods of Rosser \cite{51} or Frisch
and Wilets \cite{54} as applied to electromagnetic theory. So one need
not  bother about the requirement of electromagnetic
considerations or the necessity of a balance between the forces of
electromagnetic and gravitational origins, in invoking the
magnetic-type component in gravity. For another alternative and interesting approach to Maxwellian Gravity, Bergstr$\ddot{o}$m's \cite {56}  approach to the origin  of  magnetic field  is worth noting. In this approach one can describe the gravitomagnetic force as a Coriolis force resulting from Thomas rotation caused by the gravitational force. This point was transparently clear to Borgstr$\ddot{o}$m.\\ 
 
Further it is to be noted that even without the aid of special relativity ,one can infer the gravitational analogues of Maxwell-Lorentz equations by combining three ingredients,viz., \\
(i) the laws of gravitostatics;\\ 
(ii) the Galileo-Newton principle of relativity ( masses at rest and
masses with a common velocity viewed by a co-moving observer are
physically indistinguishable );  \\
(iii) the postulate on the existence of gravitational waves that
travel in vacuum at a speed  $\, c_{g}\, $ called the speed of
gravitational of  waves  and following  the approach of Schwinger et
al. \cite{55} to electromagnetic theory. The field equations that
emerge from this approach coincide with  the Eqs.(33)-(36) when
$\,c_{g}\,=\,c\,$. \\ 
        Judged from all these variant approaches that lead to Maxwellian Gravity, we have reasons to suspect the existing belief \cite {1} that gravitomagnetism is a manifestation of space-time curvature as described in general relativity.\\
\section{Consequences of the Maxwellian Gravity}
Maxwellian Gravity is very much analogous to Maxwell's electromagnetic
theory as revealed by the form of its equations. Therefore gravitational
phenomena very much analogous to those of electromagnetic theory are
not surprising to be revealed by this theory. However few concepts and
results of unconventional nature and importance may be discussed as
under.  \\ 
        
Let us start with the concept of field energy density in gravitation theory.This concept is of some historical as well as physical importance in any field theory of gravity.It has been pointed out by McDonald that\cite {28}: \begin{quote}`` J. C. Maxwell ended his great paper 1864 `A Dynamical Theory of Electromagnetic Field' with remarks on Newtonian gravity as a vector field theory.He was dissatisfied with his results because the potential energy of a static configuration is always negative but he felt this should be re-expressible as an integral over field energy density which, being the square of the gravitational field, is positive''.\end{quote} 
 However, this dissatisfaction of  Maxwell over a vector theory of
 gravity  can be overcome if energy density of gravito-electric
 (i.e.the electric-type component of gravity) and gravitomagnetic
 (i.e. the magnetic-type component ) field , respectively are defined
 with a negative sign in the following manner,viz., 
\begin{equation}
  (i)\,\,\,u_ {g\,e}\,=\,-\,{\frac{1}{2}} \epsilon_ {0\,g} \vec E_ {g} \cdot 
  \vec E_ {g}\,\,\,\,\,\,\,(ii)\,\,\,u_ {g\,m}\, = \,-\,\frac {1}{2
    \mu_ {0\,g}}\, \vec B_ {g}\cdot \vec B_ {g} 
\end{equation} 
where  $\, \epsilon_{0\,g}\,=\,{1}/{4\,\pi\,G\,}$ and
  $\,\mu_{0\,g}\,=\,4\,\pi\,G/c^{2}\,$ and the total field energy
  density is given by a sum of the above two,i.e.
\begin{equation}u_ {f\,i\,e\,l\,d}\,\,=\,\,u_ {g\,e}\,+\,u_ {g\,m} 
\end{equation} 
For a particle at rest, i.e., in gravitostatics, the only contribution
to its gravitational field energy is that due to the gravito-electric
field.This definition of gravitational field energy may most easily be
obtained by analogy with electromagnetism, noting that the negative
sign is a consequence of the attractive nature of gravity. In
gravitostatics ,it easy to compute the gravitational or gravito-electric
self energy of a sphere of radius $ R $ and rest mass $\,M_{0}\,$ with
uniform rest mass density by using Eq.(42i),which comes out as : 
\begin{equation} U_{g}\,=\,-\,{\frac{1}{2}}\,\epsilon_ {0\,g}{\int_{0}^\infty}{E_ {g}}^{2}\,4 \pi r^{2}\,dr\,\,=\,-\,\frac{3G{M_{0}}^{2}}{5R} 
\end{equation} 
The result (44) is in complete agreement with the Newtonian
result. It is to be noted that Visser \cite{57} used  exactly this
definition of gravitational field energy density in his classical
model for the electron. Such a definition of the field energy of
gravity has the  advantage of describing the correct nature of
gravitation on quantization because in analogy with electromagnetic
theory Maxwellian Gravity will eventually lead to a gauge  field  of
spin $1$ and the spin $1$ gauge fields having positive and definite field
energy, on quantization, as we know lead to a repulsive force field for identical charges of such fields. It is due to this reason Gupta \cite{58}
suggested  rejection of any spin $1$ gauge theory of gravity with field
energy being positive and definite as such fields do not account  for
the observed nature of gravitational interaction.\\ 
         
\,\,\,The law of energy-momentum conservation is one of the important
aspect of the validity of any physical theory. By analyzing Einstein's
general relativity, Denisov and Logunov \cite{59} have shown that
General Relativity does not obey this strict law of nature when matter
and gravitational field are taken together. Inasmuch  as the theories
of other physical fields, a unified conservation law of
energy-momentum exists for different forms of matter, and since there
is at present no experimental evidence of its violation (moreover, the
history of physics has always illustrated its tenacity and truth ),
there is no reason to reject this law. However,with the gravitational
Lorentz force law (30 or 37) as  revealed in this paper, the field momentum density defined by
\begin{equation}\vec N\,=\,\vec P_{g}/c^{2}\,=\,(\vec H_{g}\times\vec
  E_{g})/c^{2}
\end{equation} 
Where the gravitomagnetic field intensity  $\vec H_{g} $ is defined (in 
empty space) as
\begin{equation}\vec H_{g}\,=\,\vec B_{g}/ \mu_ {0\,g}
\end{equation}
and the gravitational Poynting vector $\,\vec P_{g}\,$ defined as
\begin{equation}\vec P_{g}\,=\,\vec H_ {g} \times \vec E_ {g}
\end{equation}
and the field energy defined by (42), it is easy to verify that
Maxwellian gravity is consistent with that sacrosanct law of
nature. The gravitational Poynting vector defined in Eq.(47) is in
agreement with that  used by Krumm and Bedford \cite {60}.\\ 
 
In empty space the fields $ {\vec E_{g}} $   and ${ \vec B_{g}}
$ of this theory satisfy the following two wave equations 
\begin{equation}\nabla^{2}\cdot\vec E_{g} -  \frac{1}{{c_{g}}^{2}}\cdot{\frac{\partial^{2}\vec E_{g}}{\partial t^{2}}}  = 0 
\end{equation}
\begin{equation}\nabla^ {2}\cdot\vec B_{g} - 
\frac{1}{{c_{g}}^{2}}\cdot\frac{\partial^{2}\vec B_{g}}{\partial t^{2}} = 0
\end{equation}
where  $ c_{g} = c $ . Thus the theory, in the spirit of Maxwell's
electromagnetic theory, predicts beyond doubt the existence of
gravitational waves traveling through empty space exactly at the speed 
of light as expected. Like electromagnetic waves these gravitational
waves are transverse in nature and carry energy momentum. Further the fields $ {\vec E_{g}} $   and ${ \vec B_{g}} $ of the new theory are derivable from potential functions
\begin{equation}\vec B_{g} = \nabla\times{\vec A_{g}},\>\>\>\>\>\>\>\>  \vec E_{g} = -\nabla \cdot \Phi_{g} - {\partial{\vec A_{g}}/{\partial t}}
\end{equation}
where ${ \Phi_{g}} $ and ${ \vec A_{g}} $ represents respectively the
gravitational scalar and vector potential of the new theory.These
potentials satisfy the inhomogeneous wave equations :
\begin{equation}\nabla^{2}\cdot\Phi_{g}\, - \,\frac{1}{{c}^{2}}\cdot\frac{\partial^{2}\Phi_{g}}{\partial t^{2}}\,=\,4\pi\,G\, \rho_ {0}\,=\,\rho_{0}/\epsilon_ {0\,g}  
\end{equation}
\begin{equation}\nabla^ {2}\cdot\vec A_{g}\, -\,
  \frac{1}{c^{2}}\cdot\frac{\partial^{2}\vec A_{g}}{\partial
    t^{2}}\,=\,\frac{4\pi\,G}{c^{2}}\vec j_{0}\,= \,\mu_{0\,g}\vec j_{0} 
\end{equation}
if the gravitational Lorenz \cite{61} gauge condition
\begin{equation}\vec{\nabla}\cdot\vec
  A_{g}\,+\,\frac{1}{c^{2}}\frac{\partial\Phi_{g}}{\partial{t}}\,=\,0
\end{equation}
is imposed. These will determine the generation of gravitational waves 
by prescribed gravitational charge and current
distributions. Particular solutions (in vacuum) are 
\begin{equation}\Phi_{g}\,(\,\vec r\,,t\,)\,=\,-\,G\,\int\,{\frac{\rho_{0}(\,\vec r^{\prime}\,,\,t^{\prime}\,)}{\,|\vec r\,-\,\vec r^{\prime}\,|}dv^{\prime}}
\end{equation} 
\begin{equation}\vec A_{g}\,(\,\vec
  r\,,t\,)\,=\,-\,\frac{G}{c^{2}}\,\int\,{\frac{\vec j_{0}(\,\vec r^{\prime}\,,\,t^{\prime}\,)}{\,|\vec r\,-\,\vec r^{\prime}\,|}dv^{\prime}}
\end{equation}
 where $\,t^{\prime}\,= \,t\,-\,{|\,\vec r\,-\,\vec
   r^{\prime}\,|}/c\,$  is the retarded time.These are called the
 retarded potentials. Thus we saw that retardation in gravity is
 possible in Minkowski space-time in the same procedure  as we adopt
 in electrodynamics.This result seems to conflict with the view \cite{62} that Newtonian gravity is entirely static, retardation is not possible until the correction due  to deviations from Minkowski space is considered.         \\ 
           
Einstein's general theory of relativity (GR) also predicts such
potentials in the weak field limit. But the vector potential and so the gravitomagnetic field of Einstein's theory differ (as we shall see below) from that of Maxwellian Gravity by a factor of 4. It is interesting to note that there is evidence for the existence of gravitational vector potential and magnetic-type component in gravity from Lunar Laser Ranging  \cite{63}  and the LAGEOS experiment \cite {64} on the
detection of the gravitomagnetic effect of the spinning Earth. General
Relativity predicts \cite{1,14,15,24} the gravitomagnetic induction field
${\vec B_{g ,GR}}$ of the spinning Earth under slow rotation
(spinning) and weak field approximation at a value given by
Eq.(3). However, in Maxwellian Gravity following the standard electromagnetic procedure of estimation of magnetic field generated by localized current distributions (see for example, Jackson \cite {65} ), the gravitomagnetic induction field of the Earth (under slow spinning motion condition when rest mass $ \simeq $ inertial mass) can be estimated at
\begin{equation}\vec B_{g}\,=\,\frac {G}{2c^{2}r^{5}}[\vec Jr^{2}\,-\,3({\vec J}{\cdot}{\vec r})\cdot{\vec r}]\,-\,\left({\frac{4{ \pi}G}{3c^{2}}}\right){\vec J}{\delta (\vec r)} 
\end{equation}
where $ \delta (\vec r )$ is Dirac's ${ \delta }$-function. Thus we
have ( in view of (3))         \\
\begin{equation}
\vec{B}_{g}\,=\,\frac{B_{g ,GR}}{4}\,-\,\left(\frac{4{\pi}G}{3c^{2}}\right)
{\vec{J}}{\delta(\vec{r})},
\end{equation}
and when $ \vec{r} \neq 0 $,
\begin{equation} 
\vec{B}_{g}\,=\,\frac{\vec{B}_{g  ,GR}}{4} 
\end{equation} 
            
So a satellite orbiting the Earth having a gravitomagnetic field presumed
at $\vec{B}_{g}\,=\,(\vec{B}_{g, GR})/4 $  would experience a
Lense-Thirring $(LT)$\cite {66} type
nodal precession at
\begin{equation}
\dot{\vec{\Omega}}_{LT, MG}\,=\,\dot{\vec{\Omega}}_{LT}/4\,=\,{ 25\%\>\>of\>\>the\>\>LT\>\>\>precession}
\end{equation} 
as the Lense-Thirring nodal precession \cite{66} is given by 
\begin{equation}
\dot{\vec{\Omega}}_{LT}\,=\,\frac{2G\vec{J}}{c^{2}a^{3}(1-e^{2})^{3/2}}
\end{equation}
where $a$ and $e$ respectively represents the semi-major axis and eccentricity of the satellite orbit.
It is to be noted that LAGEOS experiment \cite{64} measured the
Lense-Thirring precession to an accuracy of $ 20{\%}-30{\%}$ using laser
ranging to two Earth satellites as per a report by Unnikrishnan
\cite{67}. Currently we are in search of an alternative explanation for the LAGEOS result within this theory  and it  will be addressed in our future works which may include the planetary precession also. \\ 
 
Let us now come to the question of the origin of the factor of 4 in
general relativistic gravitomagnetic field. The mathematical theory of
General relativity is based on the  concept of the equality of
gravitational and inertial masses and the concept of Riemann space-time,which
are absent in Maxwellian Gravity. So the origin of the factor of 4
might be at the foundation  level. To get some insights of this
expectation, let us consider the original analysis of the thought
experiment   \cite {49}, where space-time is kept Minkowskian and by
assumption inertial mass is taken as the gravitational mass. From the
original work of the cited thought experiment, the magnitude of the gravitomagnetic component of the force between the particles turns out, in the lowest order approximation, at 
\begin{equation} F_{\cite{49}}\,\simeq\,\frac
  {2G\,m_{01}\,m_{02}\,v^{2}}{\,c^{2}\,r^{3}}\,\end{equation}
But in Maxwellian gravity the same force under identical situation
comes out as 
\begin{equation} F_{MG}\,\simeq\,\frac
  {G\,m_{01}\,m_{02}\,v^{2}}{\,c^{2}\,r^{3}}\,\end{equation} 
Thus the co-efficients of the two forces (61)and (62) differ by a
factor of 2 and so the gravitomagnetic permeability in (61) is  twice
as implied by Maxwellian Gravity. In general relativity the
gravitomagnetic permeability is four times as implied by Maxwellian
gravity, as we have seen earlier. Since a factor of 2 originates from
the incorporation of the equivalence of gravitational and inertial
masses in the flat space-time relativistic gravity, the origin of another
factor of 2 may be ascribed to the space-time curvature.\\
          
\section {Gravitomagnetic moment, the spin and the relation between them}
The classical connection between angular momentum $\vec {L}$  and
magnetic moment $\vec{\mu_{L}}$ ( in S.I. units ) :
\begin{equation}\vec\mu_{L} = (q/2m)\vec{L},    
\end{equation}
\noindent
which holds for orbital motion even on atomic scale is
well-known. Analogously, a classical connection between angular momentum $\vec{L}$ and gravitomagnetic moment (the gravitational analog of magnetic moment) comes out in Maxwellian Gravity as

\begin{equation}\vec\mu_{gL} = (m_{0}/2m)\vec{L}
\end{equation}
\noindent
which holds for orbital motion even on atomic scale as well because of
the purely kinematic definition of gravitomagnetic moment of  a
particular volume containing proper mass currents $ \vec j_{0} = \rho_{0}\vec u(\vec{r})$  defined (in Maxwellian Gravity) by
\begin{equation}
\vec \mu_{g L} = (1/2)\int(\vec{x} \times \vec{j_0})d^{3}x =
(1/2)\int {\rho_{0}(\vec{x}\times \vec u )d^3 x}
\end{equation}
\noindent
and the standard definition of mechanical angular momentum $\vec{L} $ in terms of the velocity distribution of inertial mass densities $\rho_{m}$ :
\begin{equation}
\vec{L} = \int{\rho_{m}}(\vec{x}\times\vec{u})d^{3}x
\end{equation}
It is to be noted that $m$ in (63) and (64) represents the inertial mass of the
system (or particle) in question which is relativistically distinct from its
rest mass $m_{0}$. The notion of gravitomagnetic moment was proposed
in \cite{68} and for other studies  the reader my refer to \cite{2,3,4,5,6,7,8,9,10,67,69}.

From (63)  and (64) we can form a ratio
\begin{equation}
(\mu_{L}/\mu_{gL}) = q/m_{0} =  a \>\>\>{\rm Lorentz-invariant \ quantity}
\end{equation}
The relation (67) holds good irrespective of the magnitude of the inertial
masses.
   It is well known that the classical connection (63) fails for the intrinsic magnetic moment of electrons and other elementary particles. For electrons, the intrinsic magnetic moment is slightly more than twice as large as implied by (63), with the spin angular momentum $\vec{S}$  replacing $\vec{L}$   and the rest mass $m_{0}$ replacing the inertial mass $m$. Thus we speak of the electron having a $g$-factor of 2(1.00116) and the spin magnetic moment $\vec{\mu_{s}}$:
\begin{equation}
\vec{\mu_{s}} = (g q/2m_{0})\vec{S}\>\>\>\>{\rm (in \ S.I. \ units)}
\end{equation}
The departure of the magnetic moment from its classical  value has its
origins in relativistic and quantum-mechanical effects which will be
reiterated later in this paper in connection with the
quantum-mechanical description of spin gravitomagnetic moment.An
interesting  feature of the relation (68) is that the ratio of the
charge of interaction and the intrinsic mass $m_{o}$ appears as a
proportionality constant. If we look at the magnetic moment as the
quantity that is the source of an elementary magnetic dipole field or
as the quantity that describes, the response to an applied magnetic
torque then the proportionality constant has the structure (charge of
the field/intrinsic mass). This observation  differs from that
of Unnikrishnan \cite{67} in that the proportionality constant has the
structure (charge of the field/inertial mass). Our observation stems
from Dirac's prediction of the spin magnetic moment of electrons. This
observation allows us to advance the hypothesis that in the case of
gravitational interaction, the spin angular momentum and ``the gravitational
spin" (or the gravitomagnetic moment) are connected in a general way, with
the proportionality containing the ratio (gravitational charge (or mass)/
intrinsic mass). The proportionality constant (gravitational mass/ intrinsic
(or rest) mass) is equal to unity in the framework of Maxwellian Gravity.
Now in analogy with (68) we define the spin gravitomagnetic moment
$\vec{\mu_{gs}}$ :
\begin{equation}\vec{\mu_{gs}} = (\kappa_{s}/2)\vec{S}
\end{equation}
\noindent
with  $\kappa_{s}$ being the gravitational analog of the $g$-factor in
(68). It is interesting to note that a ratio $(\mu_{s}/\mu_{gs})$
formed in analogy with (67)  yield
\begin{equation}(\mu_{s}/\mu_{gs}) = q/m_{0}
\end{equation}
\noindent
under the condition that $g = \kappa_{s}$. Let us now make a  quantum-mechanical investigation of this condition $(g = \kappa_{s})$ using Dirac's theory of electrons in flat space-time.

Dirac's equation for the motion of a charged massive particle (say
electron) of rest mass $m_{o}$ and charge $q$ in an external
electromagnetic field having electric and magnetic components may be
written as
\begin{equation}i\hbar(\partial{\Psi}/{\partial t})\,=\,H\Psi\,=
  [c{\bf{\alpha}}\cdot({\vec P}\,-\,q\vec A_{e})\,+\,{\bf\beta}\,{m_{0}}c^{2}\,+\,q\Phi_{e}]\Psi
\end{equation}
\noindent
where  $\Phi_{e}$ and $ \vec A_{e}$  represent the scalar and vector
potentials of the external electromagnetic field, $\,\bf\alpha\,$ and
$\,\bf\beta\,$ are Dirac matrices in the representation of Bjorken and
Drell\cite {70} and other symbols have their usual meanings. If in addition to the electromagnetic field, there exists such other fields as predicted by Maxwellian Gravity, then Eq.(71) may be amended to the following generalized form
\begin{equation}i\hbar(\partial{\Psi}/{\partial t})
  =\,H\Psi\,=\,[c{\bf{\alpha}}\cdot({\vec P}\,-\,q\vec A_{e} - {m_{0}}\vec A_{g}) + {\bf\beta}{m_{0}}c^{2} + q\Phi_{e} + {m_{0}}\Phi_{g}]\Psi
\end{equation}
\noindent
where $\Phi_{g}$  and $ \vec A_{g} $ represent the gravitational
scalar and vector potentials of Maxwellian Gravity. Since there exists 
no charged particle without mass and no massive particle without
creating and yielding to gravity, the generalized Dirac equation (GDE) 
(72) is expected to describe more correctly the interaction of Dirac particles among themselves and with gravity as it takes into account the gravitational
contributions to the Hamiltonian which may not be neglected in a
variety of situations in high energy physics.

The relativistic energy of the test particle either in (25) or (26)
includes also its rest energy $  m_{o}c^{2} $. Now following the
standard procedure of obtaining Pauli's equation from (71) \cite {70},
we obtained the following generalized Pauli's equation (GPE) from the
GDE (72), in the first non-relativistic approximation : 
\begin{equation}i\hbar\left(\frac{\partial{\Psi}}{\partial t}\right) =
  {\hat{H}}\Phi{\simeq}\left[\frac{({\vec P} - q\vec{A_{e}} - m_{0}\vec{A_{g}})^{2}}{2m_{0}} + q\Phi_{e} + m_{0}\Phi_{g} - {\frac{q\hbar}{2m_{0}}}{\bf\vec\sigma}\cdot{\vec B} -{\frac{\hbar}{2}}{\bf\vec\sigma}\cdot{\vec{B_{g}}}\right]\Phi
\end{equation}                                                                 \noindent
where $ \vec B = \vec{\nabla}\times{\vec A_{e}} $,  $ \vec{B_{g}}
= \vec{\nabla} \times {\vec A_{g}} $ , $\,\,\Phi $  is the upper component of
 Dirac's  bi-spinor  $ \Psi = \left( \begin{array}{c}{\Phi} \\ {\chi}
  \end{array} \right) $ and other symbols have their usual
meanings. The last two terms in the non-relativistic Hamiltonian (73 ) have the form of the potential energy of a dipole in external fields. Thus in the first approximation, the charged particle behaves as a particle having a spin magnetic moment
\begin{equation}\vec{\mu_{s}} = \left (\frac{q\hbar}{2m_{0}}\right ) {\bf\vec{\sigma}} =  (q/m_{0})\vec{S}
\end{equation}
\noindent
and a spin gravitomagnetic moment
\begin{equation}\vec{\mu_{gs}} = ( 1/2 )\hbar{\bf\vec{\sigma}} = \vec{S}
\end{equation}
\noindent
where $ \vec{S} = ( 1/2 )\hbar{\bf\vec{\sigma}} $ is the spin angular
momentum of the particle in question. The relation (74) is the well
known result of Dirac's theory which when compared with (68) yields a
$g$-factor of $g$=2. The relation (75) is our new result which  when
compared with (69) yields $\kappa_{s} = 2$.

\section{Dynamics of the Dirac fermions in Maxwellian Gravity}
The dynamics of the Dirac fermions in different gravitational fields and
non-inertial reference frames was studied previously ( \cite{12,14,16,17,18} and the literature quoted therein) using different schemes. Here we  will present some results that come in the description of the interaction of a spin $ 1/2 $ particle (Dirac fermion) with the fields of Maxwellian Gravity adopting the very procedures of the standard electromagnetic case. The correct description of the interaction of a spin  $ 1/2 $ particle with an
electromagnetic field is given by the effective Hamiltonian \cite{71}
derived from the Dirac equation (71) using the method of
Foldy-Wouthuysen (FW) transformations \cite{72}. In the gravitational case of Maxwellian gravity, the gravitational analog of the Dirac equation (71) is
\begin{equation}
i\hbar{\left(\frac{\partial \Psi}{\partial t}\right)}\,=\,[c{\bf{\alpha}}\cdot ({\vec P}\,-\,m_{0}\vec A_{g})\,+\,{\bf\beta }m_{0}c^{2}\,+\,m_{0}\Phi_{g}]\Psi
\end{equation}
\noindent              
and the corresponding effective Hamiltonian can analogously be obtained as 

\begin{equation} 
H_{eff}=\,\frac{(\vec P - m_{0}\vec A_{g})^{2}}{2m_{0}}\,-\,
\frac{{\vec P}^{4}}{8{m_{0}}^{3}c^{2}}\,+\, m_{0}\Phi_{g}\,-\,
\frac{\hbar}{2}{\vec{\sigma}}\cdot{\vec{B_{g}}}\, +\,
\frac{{\hbar}^2[{\vec\nabla}^2
\Phi_{g}]} 
{8m_{0}c^2}\,+\,\left
(\frac{\hbar}{4m_{0}c^{2}r}\right)
\left(\frac{d\Phi_{g}}{dr}\right)
{\vec\sigma}\cdot{\vec{L}} 
\end{equation} 

\noindent
where the various terms have their respective gravitational meanings
corresponding to their electromagnetic counter parts. Hence,
what we may call {\em{gravitational fine structure}} comes from three
terms: \\ 
       
$\bullet \> \>\>  -\,\frac{{\vec P\,}^{4}}{8{m_{0}}^{3}c^{2}}\>\> \> \> \> $ 
$\>\>\>\>\>\>\>\>\>\>\>\>\>\>${\em { relativistic mass increase}}\,;  \\ 
     
$ \bullet \> \>\>  \frac{\hbar^2 [\nabla^2 \Phi_g]}{8 m_0 c^2}\,=\,-\,\frac{\hbar^2 [\vec{\nabla}\cdot \vec{E_{g}}]}{8m_0 c^2} \> \> \>\>\>$ $\>\>\>\>\>\>$ {\em{ Gravito-Darwin term}}  \\ 
  
which clearly admits a physical interpretation similar to that of the usual
electromagnetic Darwin term, reflecting the {\em  Zitterbewegung } fluctuation of the fermion's position, that make the fermion sensitive to the average gravitational potential in the vicinity of its average position. Gravitational Darwin term has already been discussed earlier \cite{12,16} in the description of the behavior of Dirac particles immersed in the fields of General Relativity  where the gravitational Darwin term differs from the present term by a factor of 4 as expected  in view of the analysis made in this paper.Hehl and Ni \cite{73} ,who derived the inertial effects for a Dirac particle in accelerated and
rotating frames using Minkowski space-time, have also found a term ( what they call red shift to kinetic energy) that has been shown \cite{12} to have the form of a Darwin term. Interestingly, Hehl and Ni's red shift to kinetic energy term - interpreted in  \cite{12} as the gravitational analogue of Darwin term - is four times larger than the Gravito-Darwin term we found here.This difference may be interpreted as a manifestation of  the violation of the equivalence principle.    \\ 
      
$\bullet\> \> \>\>\> \left(\frac{\hbar}{4m_{0}c^{2}r}\right )\left(\frac{d\Phi_{g}}{dr}\right){\bf{\vec\sigma}}\cdot{\vec{L}} =\,\left(\frac{1}{2m_{0}c^{2}r}\right)\left(\frac{d\Phi_{g}}{dr}\right){\vec{S}}\cdot{\vec{L}}\>
\> \> \> \>\>\>\>\>\>\>\>\>$  {\em gravitational spin-orbit term.}\\ 

This spin-orbit term is due to the interaction of the fermion's
gravitomagnetic moment (i.e. the spin) with the gravitomagnetic field
it sees due to its motion and automatically includes the Thomas
precession as in the electronic case. For recent discussions on
gravitational spin-orbit coupling the reader may refer to
\cite{12,16,17} where the gravitational spin-orbit term is shown as the EEP( Einstein's equivalence principle)
violating term .But the inertial spin-orbit coupling term,which first turned up as a result of Hehl and Ni's calculation \cite{73} coincides with the prediction here under the condition :
inertial acceleration $\,\vec a\,=\,-\,\vec E_{g}\,$. Hence we 
observe this term as not the EEP-violating term.\\ 
         
   Corresponding to the full Zeeman effect (Dirac) in the electronic case, we
here predict what we call the Gravito-Zeeman effect as arising out of two
terms : the orbital part is
$ -\frac{1}{2}({\vec P}\cdot{\vec A_{g}} + {\vec A_{g}}\cdot{\vec P})
= -\frac{1}{2}\vec B_{g}\cdot{\vec {L}}$, 
\noindent
where $ \vec A_{g} = -\frac{1}{2}({\vec{r}}\times{\vec B_{g}})$ for a
weak uniform gravitomagnetic field $\vec B_{g} $ and the
spin part is  $-\frac{\hbar}{2}{\bf{\vec\sigma}}\cdot \vec{B_{g}}$. Combining the two parts we get
\begin{equation} 
H_{gravito-Zeeman} = -\frac{1}{2}{\vec B_{g}}\cdot({\vec{L}} + 2\vec{S})
\end{equation}
\noindent
where $\vec{S} = \frac{\hbar}{2}{\bf{\vec\sigma}}$. Note the factor of $2$
for the fermion's intrinsic gravito-gyro-magnetic ratio.\\
 From the GDE  (72) or from the GPE (73) one can obtain the
 gravitationally modified (or generalized ) Zeeman effect (GZE), in the same 
 procedure as in the electromagnetic case, corresponding to the Hamiltonian : 
 \begin{equation}H_{GZE}\, =\, -\,\frac{1}{2}\left({\vec B_{g}}\,+\,\frac{q\vec B\,}{m_{0}}\right)\cdot({\vec{L}} + 2\vec{S})
\end{equation} 
In specific situations where $\vec B_{g}\,+\,q\vec B/m_{0}\,=\,0\,$,
Zeeman effect can be nullified. In other cases Zeeman effect gets
modified  in presence of a gravitomagnetic field. Hence, Zeeman Effect
may be employed for the detection of gravitomagnetic of Earth or other
rapidly rotating astrophysical objects. Another interesting inference
of this analysis is that unlike the spin  magnetic moment, the spin
gravitomagnetic is independent of the charge  and mass of the Dirac
particle. Therefore charged Dirac particles that can be easily handled electromagnetically in any desired way to move with relativistic
speeds in the gravitational field of the earth, can also be employed
to detect the gravitomagnetic field of the earth. Further Eq.(79) may be rewritten as
\begin{equation}H_{GZE}\, =\, -\,\frac{1}{2}\left({\vec B_{g}}\,+\,\vec {\omega_{c}}\right)\cdot({\vec{L}} + 2\vec{S})
\end{equation} 
where $\,\vec\omega_{c}\,=\,q\vec B/m_{0}\,$ is the cyclotron
frequency. Hence here seems a possibility of utilizing the cyclotrons
for the detection of the gravitomagnetic field of the Earth.
\section{Concluding Remarks}
For a massive charged Dirac $(spin{\frac{1}{2}})$ fermion we found (in
Sec.3) that the gravitomagnetic moment (= the spin) is independent of
its rest mass and electric charge. Hence we remark that all spin
$\frac{1}{2}$ particles possessing whatever rest masses and electric
charges must interact with gravity identically under identical
conditions. In this sense the spin-gravity and spin-spin interaction
is a universal phenomenon for all spin $\frac{1}{2}$ particles
irrespective of their rest masses and electric charges.The spin-spin
interaction Hamiltonian of two Dirac particles having spins $\vec
S_{1}$ and $\vec S_{2}$  here comes out( in analogy with the electromagnetic
case) as :
\begin{equation}H_{spin-spin}\,= -\,\vec B_{g\,\vec S_{1}}\cdot\vec S_{2}\,=\,\left({\frac{8{ \pi}G}{3c^{2}}}\right){\vec S_{1}\cdot\vec S_{2}}{\delta (\vec r)}\,-\,\frac {G}{c^{2}r^{5}}[(\vec S_{1}\cdot\vec S_{2})r^{2} - 3 ({\vec S_{1}}{\cdot}{\vec r})\cdot({\vec S_{2}}{\cdot}{\vec r}) ] 
\end{equation}
where $ \delta (\vec r )$ is Dirac's ${ \delta }$-function.It must be
noted that this is different from Lense-Thirring effect which involves 
the rotation of the frame of the reference.The $\,\delta$-function
contribution to the Hamiltonian in (81) is the gravitational analogue
of hyperfine interaction as it resembles the interaction of the spin
of the electron with the magnetic field of the nucleus of an atom.This 
effect is very negligible in atomic cases because the coefficient of
this contribution contains the ratio of $G$ and $c^{2}$ . When $\,r\neq\,0\,$
we have
\begin{equation}H_{spin-spin}\,= -\,\vec B_{g\,\vec S_{1}}\cdot\vec
  S_{2}\,=\,-\,\frac {G}{c^{2}r^{5}}[(\vec S_{1}\cdot\vec S_{2})r^{2} - 3 ({\vec S_{1}}{\cdot}{\vec r})\cdot({\vec S_{2}}{\cdot}{\vec r}) ] 
\end{equation}
and this form of interaction is encountered in torsion gravity
\cite{20,46}. So the origin of the torsion field seems to have a
link with the spin.\\ 
         
Now coming to the macroscopic situation, we found Dirac particles 
coupling to the gravitomagnetic field of the the Earth will have the
interaction Hamiltonian
\begin{equation}
H_{max. grav.} = -{\vec{S}}\cdot{\vec B_{g}}
\end{equation}
where $\vec B_{g}$ is given either by (56)or (58) in the framework of Maxwellian gravity, while the interaction Hamiltonian in the framework of general
relativity is considered \cite{24} as
\begin{equation} 
H_{GR} = -{\vec{S}}\cdot{\vec B_{g ,GR}}                                
\end{equation}
\noindent

where $\vec B_{g ,GR}$ is given by (3). Since $ {\vec B_{g}} \neq
{\vec B_{g ,GR}}$ as we have seen , the two
  Hamiltonians (83) and (84) will be different and this difference
  will manifest itself in experimental tests of quantum gravity phenomena.
  In this connection the novel experimental proposals put forwaded by  Camacho
  \cite{15,74} recently for quantum-mechanical detection of the
  gravitomagnetic field of the Earth are noteworthy.\\
         
 Maxwellian Gravity is a relativistic vector model theory of
gravity in flat space-time  and the theory having the classical Newtonian limit is in conformity with the correspondence principle and the weak equivalence
principle (WEP) at the non-relativistic level. Unlike Maxwell's
equations of electromagnetic theory which represent mathematical
expressions of certain experimental results, the field equations of
Maxwellian Gravity represent mathematical expressions of certain
theoretical deductions (without any additional postulation) from other 
established theories, the applicability of which to any physical situation needs verification through extensive theoretical as well as experimental
work.The laws of Maxwellian gravity are valid in the world of inertial
frames, while general relativity describes physics in the world of
non-inertial frames. Since general relativity provides for the
existence of inertial frames in the world of non-inertial frames
\cite{75}, this theory is expected to find application in the area common to
both general and special relativity. It is to be noted that the so
called very strong Equivalence Principle \cite{1} does not state what
are the special relativistic laws of gravity although it states that
for every point-like event of space-time, there exists a sufficiently
small neighborhood such that in every local, freely falling frame in
that neighborhood, all laws of physics obey  the laws of special
relativity. Maxwellian Gravity is what represents the laws of special
relativistic gravity. Although originally developed in
flat-space-time,the curved space-time version of this theory is not
difficult to achieve. But the present treatment of relativistic gravity in a flat space-time has two great advantages. Firstly, it provides us with a more  uniform description of the gravitational field and electromagnetic
fields. Secondly, it may enable us to carry out the quantization of
gravitational field of this type by following the same procedure as we 
use for the electromagnetic field. On quantization ,Maxwellian gravity
with its field energy being negative and definite  is expected to
correspond to gravitational quanta or gravitons of vanishing rest mass
and spin 1 which may produce the observed attractive type
gravitational field for like charges of such fields. So from a practical point of view the theory
presented here may be very useful particularly in respect of its
quantization. Further study of the Generalized Dirac
Equation, viz.Eq.(72), suggested in this work may lead to new insights.\\ 
        
As regards the ``three or four crucial tests'' \cite {1,29,30}  of general relativity we are close to get an alternative explanation for the so called Non-Newtonian excess precession of Mercury and other planets  within the framework of Maxwellian Gravity. This problem will be addressed  separately in our
future works.\\ 
         
\textbf{Acknowledgments} :
The authors thank Prof. N. Barik, Dept. of Phys. Utkal
University, Bhubaneswar and N.K. Behera, Dept. of Chem., U.N. College, Soro,
Balasore for encouragements, fruitful discussions and suggestions. The
authors also  acknowledge the help received from the Institute of
Physics, Bhubaneswar for using its Library and Computer Centre for this work.\\

                                                                                                                                                              %\textbf{References}

\end{document}